\documentclass[aps,pre,onecolumn,showpacs,a4paper]{revtex4}
\usepackage{amssymb,amsmath}
\usepackage{epsfig}
\usepackage{graphicx}
\usepackage{url}

\begin{document}

\title{The parallel TASEP, fixed particle number and weighted Motzkin paths}

\author{Marko Woelki\footnote[1]{marko.woelki@DLR.de}}

\affiliation {Institute of Transportation Systems, German Aerospace Center, Rutherfordstra{\ss}e 2, 12489 Berlin, Germany}

\begin{abstract}
In this paper the totally asymmetric exclusion process (TASEP) with parallel update on an open lattice of size $L$ is considered in the maximum-current region. A formal expression for the generating function for the weight of configurations with $N$ particles is given. Further an interpretation in terms of $(u,l,d)$-colored weighted Motzkin paths is presented. Using previous results (Woelki and Schreckenberg 2009 {\it J. Stat. Mech} P05014, Woelki 2010 {\it Cellular Automata}, pp 637-645) the generating function is compared with the one for a possible 2nd-class particle dynamics for the parallel TASEP. It is shown that both become physically equivalent in the thermodynamic limit.
\end{abstract}

\maketitle

\section{Introduction}
One of the very rare exactly solvable non-equilibrium systems is
the totally asymmetric simple exclusion process (TASEP), see
\cite{Mallick_rev} and references therein. The model is
defined on a 1d discrete lattice with $L$ sites that are either
occupied by a single particle or empty. In the latter case they
can be thought of as occupied by a single hole. In its version with
parallel update the TASEP is equivalent to a special case of the Nagel-Schreckenberg model
for traffic flow \cite{NaSch}. Here all bonds are updated simultaneously while in the original
formulation, the random-sequential update, only one particle move can occur per
infinitesimal time-step.
In general, particles enter the lattice on site $1$ with probability $\alpha$
and leave the lattice from site $L$ with probability $\beta$.
In between particles on sites $l$ ($l=1,2,\dots, L-1$) may move with probability $p$
if the target site $l+1$ is empty. For later use we introduce also the symbol $q=1-p$
which denotes the probability that any of the possible moves is not executed.
Note that for random-sequential update those
probabilities are replaced by rates. In the thermodynamic limit $L\rightarrow \infty$
there are three different phases: a low-density phase, a high-density
phase and a maximum-current phase. Here, the maximum-current
phase is special, since all physical quantities become independent of the input and
output probabilities $\alpha$ and $\beta$ -- the system behaves as if
$\alpha=\beta=p$. The average density in the maximum-current phase is $1/2$.
In the limit $p\rightarrow 0$ only one move per time-step
occurs and one recovers the well-studied TASEP in continuous time.\\
While for random updating many results are known there is a definite lack of results for
the parallel updating scheme \cite{Blythe}. One reason is that the structure of the exact solution
\cite{ERS}, \cite{DeGier} appeared to be much more involved. Based on a simplified
formulation of the exact solution \cite{woelki_par} the present article tries to close
this gap a little more since especially the parallel update is important for practical
modeling of traffic \cite{NaSch}. In this article we calculate as the main result
the weight for $N$ particles and $M$ holes in the open system.
For continuous time this quantitiy has been
calculated by Derrida et. al. twenty years ago in \cite{Derrida_2class}.
It further has a natural interpretation as the normalization of a related system
on a ring with one second-class particle. This normalization is given by so-called
Narayana numbers \cite{Blythe_dyck}. We will derive the analogous results for the parallel case here.
Recently a connection between equilibrium lattice walks and the normalization of the TASEP in different variants
has been established. One of those interpretations involves so-called Motzkin paths.
A Motzkin path is a path defined on the triangular lattice. It starts at
$(0,0)$ and ends at $(n,0)$ with never going below the horizontal axis.
Possible are the steps $\{U,L,D\}$, where $U=(1,1)$ is an up-step, $L=(1,0)$
is a level step and $D=(1,-1)$ is a down step. Let $M(n)$ denote the
set of those Motzkin paths. Then their number $|M(n)|$ is given by the
$n$-th Motzkin number. The problem can also be formulated as a random walk
in 1D that starts and ends at site 0 (or 1). If the horizontal axis represents time, the two-dimensional
walk corresponds to the trajectory of the walker. A level step means that the
walker has not moved during the time-step. An up-step (down-step) means that the
walker increases (lowers) its coordinate by one. A Brownian excursion \cite{Derrida_2class},
\cite{Derrida_Brownian} is a special Motzkin path that never even touches the horizontal axis apart from the first and last vertex.
It can simply be constructed from a Motzkin path by adding an up-step at the beginning
and a down-step at the end \cite{Blythe_dyck}. In this paper we will find a Motzkin path interpretation
of the generating function for $N$ particles on an open lattice of $L$ sites. For this interpretation we use
a coloring of those different steps as it was considered in \cite{Schork}.\\
The paper is organized as follows. First, we consider the generating function of the
open TASEP and show how it is related to the generating function of weighted Motzkin paths. After taking
the thermodynamic limit we obtain expressions for the generating function of weighted Motzkin paths
at given length. We calculate the TASEP-generating function for the weight of configurations with
given number of particles. Then we see how this function is interpreted in terms
of a coloring of the corresponding Motzkin path. We write an expression for the weight that
the open TASEP system contains $N$ particles. Finally we introduce a second-class particle dynamics on the ring
that has conserving dynamics and relate it to the one obtained for the TASEP in the thermodynamic limit.
\section{The TASEP and the matrix ansatz}
The matrix representation that Derrida et. al. \cite{Derrida_matrix} presented for the known recursion relations of the TASEP came as a very elegant and compact formulation of a non-equilibrium steady state. Since then, this technique has become very successful to calculate many stationary properties for the TASEP and related models, see \cite{Blythe} for a recent  review. The authors of \cite{Derrida_matrix} found that the stationary weight for a lattice configuration $(\tau_1, \dots, \tau_L)$ of the TASEP with random-sequential update can be written as
$F(\tau_1,\dots, \tau_{L})=\langle W| \prod_{i} (\tau_i
D+(1-\tau_i)E )|V\rangle.$
In this notation the matrix $D$ represents occupied sites ($\tau=1$) and the matrix $E$ represents empty sites $\tau=0$. The boundary vectors $\langle W|$ and $|V\rangle$ ensure that the resulting matrix product is reduced to a scalar $F$. Those weights $F$ are stationary if the matrices and vectors involved satisfy the set of relations $DE=D+E$, $\langle W|E=\alpha^{-1}\langle W|$ and $D|V\rangle=\beta^{-1}|V\rangle$, now widely known as the DEHP-algebra. The matrices and vectors are half-infinite and there are different representations that have certain combinatorial interpretations \cite{Brak}. Note that $\langle 1|(D+E)^L|1 \rangle$ is given by Catalan numbers. It was shown in \cite{Derrida_2class} for the case of a single second-class particle on the ring that the weight for a fixed number of particles $N$ and $L-N$ holes is given by
\begin{equation}
\label{ra}
[x^Ny^{L-N}]\langle 1|(xD+yE)^{L}|1 \rangle = \frac{1}{L+1}{L+1 \choose N}{L+1 \choose N+1}.
\end{equation}
Here $x$ and $y$ are the fugacities for particles and holes respectively. The combinatorial term on the right-hand side is known as the Narayana numbers \cite{Blythe_dyck}. Among many other applications they are known to count Dyck paths -- those are Motzkin paths without level steps. In \cite{Derrida_Brownian} this formula (or more precisely the equivalent formula of \cite{Mallick}) was used for the TASEP with open boundaries, too. The reason why in both models the weights for $N$ particles are the same is that the DEHP algebra holds for both models under the same Ansatz.
Now to the parallel-update TASEP. The exact stationary state is known to be
of a matrix form, too \cite{ERS,DeGier}. From the findings in \cite{woelki_par} it follows that
the weight of a configuration in the maximum-current phase can be written as
\begin{equation}
\label{ans}
F(\tau_1,\dots, \tau_{L})= q^{-W(\{\tau\})}\langle W| \prod_{i}( \tau_i
D+(1-\tau_i)E) |V\rangle
\end{equation}
with $q=1-p$. In this paper we interpret $W(\{\tau\})$ as the number of particle-hole domain walls in the configuration $\{\tau\}$. This is one possible choice
following \cite{woelki_par}. Again the vectors $\langle W|$ and $|V\rangle$ reflect the boundaries and the matrices $D$ and $E$ represent particles and holes respectively but are different from those for the random-sequential case.
The operators obey the relations found by Evans et. al. in \cite{ERS}
\begin{eqnarray}
\label{WE}
\langle W|E &=& q \langle W|,\\
\label{DE}
DE &=& q\left[D+E+p \right],\\
\label{DV}
D|V\rangle &=& q |V\rangle
\end{eqnarray}
which generalize the DEHP algebra. It was shown in \cite{ERS} that the normalization can be expressed as
\begin{equation}
\label{norm}
Z_L=z_L+pz_{L-1}, \quad{\rm with}\; z_L=\langle 1|(D+E+p)^L|1 \rangle = \sum\limits_{t=0}^{L}\frac{1}{L+1}{L+1 \choose t}{L+1 \choose t+1}q^t.
\end{equation}
In contrast to (\ref{ra}) the term on the right-hand side is the $(L+1)$th Narayana polynomial.
Note that a possible representation is
\begin{eqnarray}
\label{repres}
E=\left(\begin{array} {rrrrr} q & 0 & 0 & 0 & \dots\\ \sqrt{q} & q & 0 & 0 & \dots\\ 0 & \sqrt{q} & q & 0 & \dots\\ 0 & 0 & \sqrt{q} & q & \dots\\ \dots & \dots & \dots & \dots & \dots \end{array} \right),\;
D=\left(\begin{array} {rrrrr} q & \sqrt{q} & 0 & 0 & \dots\\ 0 & q & \sqrt{q} & 0 & \dots\\ 0 & 0 & q & \sqrt{q} & \dots\\ 0 & 0 & 0 & q & \dots\\ \dots & \dots & \dots & \dots & \dots \end{array} \right)
\end{eqnarray}
along with $\langle W|=\langle 1|$ and $|V\rangle =|1 \rangle$. This representation is a direct generalization of one known representation for the continuous-time case $q\rightarrow 1$ \cite{DeGier}.
\section{The TASEP and fixed number of particles}
To calculate the weight for $N$ particles on a lattice of size $L$, our aim is to study $\mathcal{Z}_L(x,y)=\sum_{\{\tau\}} F_L(\{\tau\})\delta_{\sum\tau,N}x^Ny^{L-N}$.
For details of the following steps, the reader is referred to Appendix \ref{prelim}. First, we note (\ref{ZS}) that $\mathcal{Z}_L(x,y)$ is given by
\begin{equation}
\label{recm}
\mathcal{Z}_L(x,y)=m_L( x,y) + p(x+y)m_{L-1}(x,y) + p^2xym_{L-2}(x,y).
\end{equation}
Our intention is to study the generating function
\begin{equation}
\label{Mxy}
\mathcal{M}(x,y,\lambda,\mu) :=\sum\limits_{L} m_L(x,y,\mu) \lambda^L = \sum_{n\geq 0}(\lambda\mu)^n
\langle 1|\left(xD+yE+\frac{p}{q}xy\lambda DE\right)^n |1 \rangle =: \sum_{n\geq 0}\langle 1|\tilde{C}^n|1 \rangle \mu^n
\end{equation}
which follows from (\ref{ab}). The formula for the generating function shows its two 'faces'. On the one hand $\mathcal{M}$ has an interpretation as contributing to the grand-canonical normalization $\sum_L\mathcal{Z}_L\lambda^L$ for TASEPs of size $L$ (in terms of the TASEP fugacity $\lambda$) and on the other hand $\mathcal{M}$ will be interpreted as the generating function of Motzkin paths of length $n$ with fugacity $\mu$. We could also take $\lambda=1$ since the exponent of $\lambda$ can already be worked out by replacing $x\rightarrow x\lambda$ and $y\rightarrow y\lambda$. However at this stage it will turn out useful.
\subsection{Motzkin path interpretation}
The use of the explicit representation (\ref{repres}) yields the tri-diagonal matrix
\begin{eqnarray}
\label{tildeC}
     \tilde{C} = \left(\begin{array} {lllll} w_l(x,y,\lambda) & w_u(x,y,\lambda) & 0 & 0 & \dots\\ w_d(x,y,\lambda) & w_l(x,y,\lambda) & w_u(x,y,\lambda) & 0 & \dots\\ 0 & w_d(x,y,\lambda) & w_l(x,y,\lambda) & w_u(x,y,\lambda) & \dots\\
     \dots & \dots & \dots & \dots & \dots \end{array} \right).
\end{eqnarray}
This is the actual starting point of the problem with (\ref{Mxy}) being interpreted in terms of a weighted Motzkin path with the transition matrix (\ref{tildeC}) with $w_l$, $w_u$ and $w_d$ being the weights for level-steps, up-steps and down-steps, respectively. Related tri-diagonal matrices appear in the TASEP with a second-class particle \cite{Derrida_2class} and in the partially asymmetric exclusion process \cite{Blythe_PASEP} for example.
For the first element of the $n$-th power of this matrix we find due to the tri-diagonal structure \cite{Derrida_2class}
\begin{equation}
\label{mn}
\langle 1 | \tilde{C}^n | 1 \rangle = \sum\limits_{r=0}^{[n/2]}\frac{1}{r+1}{n \choose 2r}{2r \choose r}w_l^{n-2r}w_u^rw_d^r.
\end{equation}
with
\begin{equation}
\label{lud}
w_l(x,y,\lambda)= q(x+y)\lambda+pxy\lambda^2(1+q),
\quad\quad w_u(x,y,\lambda)=\sqrt{q}x\lambda+pxy\lambda^2\sqrt{q}, \quad\quad w_d(x,y,\lambda)=\sqrt{q}y\lambda+pxy\lambda^2\sqrt{q}.
\end{equation}
Note that for random update (\ref{mn}) turns into the well-known formula of \cite{Derrida_2class} and can be extracted to (\ref{ra}). Hence (\ref{mn}) is the total weight for all Motzkin paths of length $n$ with weights (\ref{lud}). From the first-return decomposition one finds \cite{Schork} that $\mathcal{M}$ fulfills $\mathcal{M}=1+w_l(x,y,\lambda)\mu\mathcal{M}+w_u(x,y,\lambda)w_d(x,y,\lambda)\mu^2\mathcal{M}^2$. We see that the polynomials $w_u(x,y,\lambda)$ and $w_d(x,y,\lambda)$ only appear as a product $w_uw_d$. This is because a path from height $0$ to $0$ has the same number of up- and down-steps. Hence up-/down-step pairs have weight $w_{ud}:=w_u w_d$ and the generating function reads \cite{Derrida_2class}, \cite{Schork} in terms of the fugacity $\mu$ counting Motzkin steps:
\begin{equation}
\label{Mxy2}
\mathcal{M}(w_l,w_{ud},\mu)=\frac{1-w_l\mu-\sqrt{(1-w_l\mu)^2-4w_{ud}\mu^2}}{2w_{ud}\mu^2}.
\end{equation}
In Appendix \ref{Eq}, $\mathcal{M}$ is calculated alternatively, following calculations in \cite{woelki_par}. This is done without use of an explicit representation just by manipulation of the general matrix $\tilde{C}$ in (\ref{Mxy}) to calculate $(1-\tilde{C})^{-1}$ and finally taking the matrix element. The resulting expression which involves a function that appears in \cite{Blythe_dyck} for several lattice paths is shown to agree with (\ref{Mxy2}).\\
The radius of convergence of $\mathcal{M}(\mu)=\sum_{n\geq 0}\langle 1|\tilde{C}^n|1 \rangle \mu^n$ reads with (\ref{Mxy2}) $\mu_*=(w_l + 2\sqrt{w_{ud}})^{-1}$ and the thermodynamic contribution of $\langle 1|\tilde{C}^n|1 \rangle$ can be calculated by standard techniques \cite{Blythe}. Here, we find
 \begin{equation}
 \langle 1|\tilde{C}^n|1 \rangle \sim \frac{(w_l+2\sqrt{w_{ud}})^{n+3/2}}{2\sqrt{\pi}n^{3/2}w_{ud}^{3/4}}.
 \end{equation}
 The average number of level steps fulfills $\langle n_l \rangle = w_l\;d/dw_l\; \log m_n$. This leads for $n$ large to $\rho_l = w_l/(w_l + 2\sqrt{w_{ud}})$ and $\rho_{u} = \rho_d = \sqrt{w_{ud}}/(w_l + 2\sqrt{w_{ud}})$.
In the following we consider paths at finite length. A realization of a path of length $n$ with $r$ up-down pairs has the total weight $w_l^{n-2r}w_d^r w_u^r$. Using the explicit expressions (\ref{lud}) and expanding leads to the weight
\begin{equation}
w_l^{n-2r}w_d^r w_u^r = \sum\limits_{s=0}^{r}\sum\limits_{t=0}^{r}\sum\limits_{c=0}^{n-2r}{n-2r \choose c}(1+q)^{c}(qx+qy)^{n-2r-c}q^r{r\choose s}x^{r-s}{r \choose t} y^{r-t} \times (pxy)^{c+s+t}\lambda^{n+c+s+t}.
\end{equation}
Now working out the power of $pxy$ and inserting into (\ref{mn})
leads
 \begin{table}
\begin{center}
\begin{tabular}{lll}
\begin{tabular}{c|c|c}
Red Step & Weight & Total number\\\hline
$U_r$ & $\sqrt{q}x$ & $r-s$\\
$D_r$ & $\sqrt{q}y$ & $r-t$\\
$L_r$ & $q(x+y)$ & $L-2B-(2r-s-t)$
\end{tabular}
& \quad\quad\quad\quad &
\begin{tabular}{c|c|c}
Black Step & Weight & Total number\\\hline
$U_b$ & $\sqrt{q}$ & $s$\\
$D_b$ & $\sqrt{q}$ & $t$\\
$L_b$ & $(1+q)$ & $B-s-t$
\end{tabular}
\end{tabular}
\end{center}
\caption{Enumeration of the (2,2,2)-colored Motzkin path with red (left table) and black steps (right table).}
\label{tab}
\end{table}
by comparison with (\ref{ab}) to $m_n=\sum_{c\geq 0}(pxy)^c f_{n,c} \lambda^{n+c}$ with
\begin{equation}
\label{fab}
f_{n,c}= \sum\limits_{r=0}^{[n/2]}\frac{1}{r+1}{n \choose 2r}{2r \choose r}\sum\limits_{s=0}^{r}\sum\limits_{t=0}^{r}{n-2r \choose c-s-t}{r\choose s}{r \choose t}(1+q)^{c-s-t}(qx+qy)^{n-2r-c+s+t}q^r x^{r-s} y^{r-t}.
\end{equation}
\subsection{Application to the TASEP}
To apply this formula to the TASEP, we first remark that the TASEP system size $L$ obviously enters as $L=n+c$. The result for the TASEP then reads (\ref{mncompl})
\begin{equation}
\label{mL}
m_L=\sum\limits_{B\geq 0}(pxy)^B \mu^{L-B} f_{L-B,B},
\end{equation}
where the $f_{L-B,B}$ are
\begin{equation}
\label{f}
f_{L-B,B}= \sum\limits_{r=0}^{(L-B)/2}\frac{1}{r+1}{L-B \choose 2r}{2r \choose r}\sum\limits_{s,t=0}^{r}{L-B-2r \choose B-s-t}{r\choose s}{r \choose t}(1+q)^{B-s-t}(qx+qy)^{L-2B-2r+s+t}q^r x^{r-s} y^{r-t}.
\end{equation}
What we actually did in working out (\ref{f}) is dividing the three weights $w_l(x,y,\lambda)$, $w_u(x,y,\lambda)$ and $w_d(x,y,\lambda)$ in two terms each. $w_l$ was split in $q(x+y)\lambda$ and $(1+q)pxy\lambda^2$, $w_u$ in $\sqrt{q}x\lambda$ and $\sqrt{q}pxy\lambda^2$ and $w_d$ equivalently. The following section shows how that appears naturally in coloring  Motzkin paths.

\subsection{Coloring of the Motzkin path}
\begin{figure}
\centering
\includegraphics[height=1.4cm]{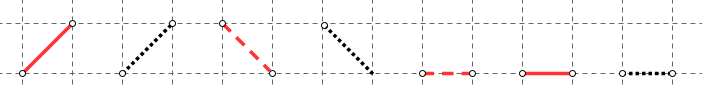}
\caption{The possible steps in the (2,3,2)-colored Motzkin path.}
\label{allowed}
\end{figure}

A $(u,l,d)$-colored Motzkin path \cite{Schork} has up-steps, level-steps and down-steps in $u$, $l$ and $d$ colors respectively.  Splitting the weights (\ref{lud}) into two parts suggests to use two colors which results in a $(2,2,2)$-colored Motzkin path. We distinguish each of the two distinct steps by the colors red and black, see table \ref{tab}. $f_{L-B,B}$ is the number of (2,2,2)-colored weighted Motzkin paths with $L$ steps, $B$ of those being black steps. Further $m_L$ is the generating function of this ensemble of paths weighting each ensemble with $(pxy)^B$ (a fugacity $p$ for each black step as well as a factor $xy$). There is no need for $\lambda$ to the explanation thus we set here $\lambda=1$.\\
To actually count the paths we have to split the contribution $(x+y)$ to red level steps further: a $(2,3,2)$-colored Motzkin path is the most appropriate interpretation. We take a continuous red line $L_{r1}$ and a red dashed line $L_{r0}$ as further coloring. Fig. \ref{allowed} shows the possible steps. All red up-steps are drawn as continuous lines and all down-steps as a dashed line while black steps are drawn as dotted lines, for a better reading in black and white.
Figure \ref{motzkin232} shows an example of a (2,3,2)-colored Motzkin path of length $7$ (see figure capture). Note that the path could also touch the horizontal axis.\\
Let us consider the density of steps that are associated with weight $x$. Those are continuous red level steps, red up-steps and black steps.
We can rescale the radius of convergence $\mu_*=(w_l+2\sqrt{w_{ud}})^{-1}$ of $\mathcal{M}=\sum_n \langle 1 | \tilde{C}^n | 1 \rangle \mu^n = \sum_{n}\mu^n\sum_{b=0}^{n}(pxy)^b f_{n,b}\lambda^{n+b}$ to one, solve for $x$ and identify the physical root. Then we calculate the density of steps associated with $x$, namely $\rho = n^{-1} x\;d/dx\; \log \langle 1 | \tilde{C}^n | 1 \rangle = -x(\lambda)/\lambda/x'(\lambda)$. The resulting expression can be written as a series
\begin{equation}
\rho(\lambda) = 1-\sqrt{\frac{\lambda}{q}} + \sum\limits_{k\geq 1}\left(\frac{p\lambda}{q}\right)^k\sum\limits_{j=0}^{k}p^{k-j}\left[\frac{1}{p}{2k\choose 2j+1} - \sqrt{\frac{\lambda}{q}}{2k+1\choose 2j+1}\right].
\end{equation}
The case $\rho=1$ corresponds to $\lambda=0$ and the density decreases monotonically to $0$ where $\lambda=q^{-1}$. The density takes the value $1/2$ for $\lambda=(2-p-2\sqrt{1-p})/p^2=(1-\sqrt{q})^2/p^2$. Note that the right-hand side is equal to (\ref{zsimeq}) and equals the asymptotic value of the lattice fugacity \cite{ERS}, \cite{Blythe_dyck}.
\subsection{The weight for N particles in TASEP}
\begin{figure}
\centering
\includegraphics[height=4.5cm]{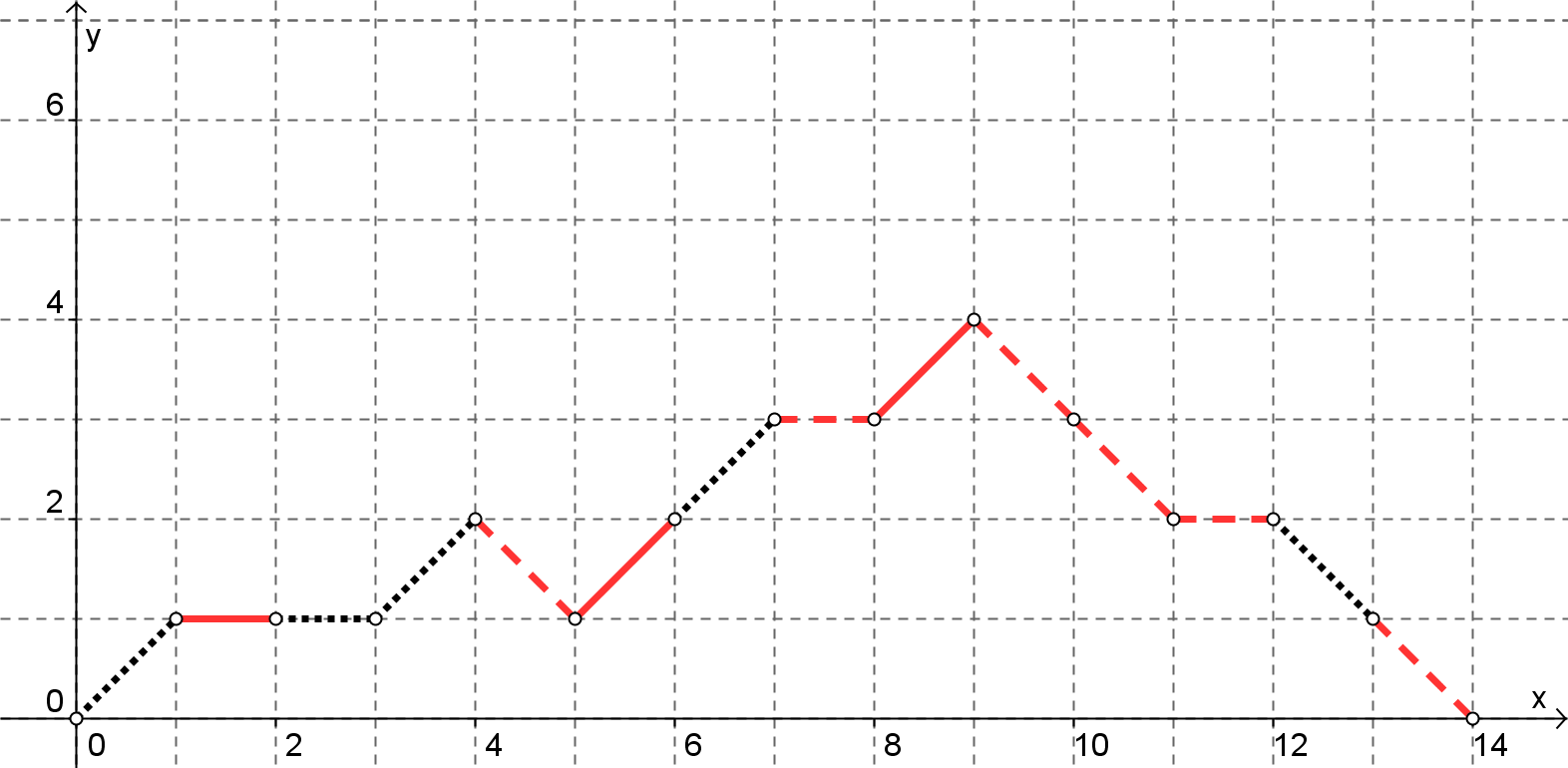}
\caption{Example of a (2,3,2)-colored Motzkin path of length $14$. The corresponding TASEP size is $19$.}
\label{motzkin232}
\end{figure}

The function $f_{L-B,B}$ can be interpreted with (\ref{mncompl}) and (\ref{ZS}) as contributing to the weight for configurations of length $L$ with at least $B$ particle-hole domain-walls. Choosing $B$ of $W$ available particle-hole domain-walls happens naturally at every update step, where a fraction $B$ of $W$ bonds adds to the flow. The $(2,3,2)$ colored Motzkin path in figure \ref{motzkin232} has $B=5$ and is one realization of a TASEP configuration with $W=6$ since the red peak at $x=9$ contributes a domain wall, too.
Note that in the TASEP the case $B=0$ corresponds to the $0$ order in $p$. Then one has $q=1$ and the continuous-time result is recovered.\\
We continue by working out the weight $Z_{M,N}$ for $N$ particles and $M$ holes in TASEP that reads due to (\ref{recm})
\begin{equation}
\label{ZMN}
Z_{M,N}=s_{M,N}+ps_{M-1 N}+ps_{M N-1}+p^2s_{M-1 N-1}.
\end{equation}
The probability that the TASEP contains $N$ particles then is Prob$(\sum_i\tau_i=N)=Z_{M,N}/Z_{M+N}$. Here, the denominator is the normalization constant (\ref{norm}).
First one shall expand $(qx+qy)^{L-2B-2r+s+t}$. The weight of a path with $B$ black steps and $r$ up-down pairs of a total of $L-B$ steps then reads
\begin{equation}
\mathcal{W}(B,L-B,r) = \sum\limits_{s,t,k}{L-B-2r \choose B-s-t}{r\choose s}{r \choose t}{L-2B-2r+s+t \choose k}(1+q)^{B-s-t}q^{L-2B-r+s+t} p^B x^{B+r-s+k} y^{L-B-r+s-k}.
\end{equation}
Now introducing the particle number $N$ and picking out the coefficient of $x^Ny^{L-N}$ yields
\begin{equation}
\{x^Ny^{L-N}\}\mathcal{W}(B,L-B,r) = \sum\limits_{s,t}{L-B-2r \choose B-s-t}{r\choose s}{r \choose t}{L-2B-2r+s+t \choose N-B-r+s}(1+q)^{B-s-t}q^{L-2B-r+s+t} p^B.
\end{equation}
Then one obtains
\begin{equation}
s_{L-N,N}=\sum\limits_{B}\sum\limits_{r=0}^{(L-B)/2}\frac{1}{r+1}{L-B \choose 2r}{2r \choose r} \{x^Ny^{L-N}\}\mathcal{W}(B,L-B,r).
\end{equation}
Note that one has due to the particle-hole symmetry that $s_{L-N,N}=s_{N,L-N}$. Further the $z_L$ entering the normalization (\ref{norm}) read $z_L=\sum_N \left(s_{L-N,N} + ps_{L-N,N-1}\right)$.

\section{Second-class particle dynamics}
In \cite{woelki_par} it was argued that second-class particle dynamics in the parallel TASEP are not as natural to define as for the generic random update. In this section a model system is presented that mimics second-class particle dynamics and for which preliminary results were published in \cite{woelki_cargo}. We change the boundary conditions and consider a closed periodic chain with one of the particles carrying a cargo. The unit of the particle and the cargo is referred to as the second-class particle. Under the parallel update every particle (with or without cargo) moves forward with probability $p$ if the target site is empty. In the most simple case the cargo is carried to the next site. However if the site behind it is occupied, the cargo can actively jump onto this particle which happens independently with probability $p$. Thus the cargo can jump to the left while its carrier moves at the same time to the right. This dynamics reminds a bit of someone standing on a train either standing still and moving with the train or actively jumping to the wagon behind.
While in the continuous-time case a 'move' corresponds to an
interchange of occupation numbers $\tau_i\tau_{i+1}\rightarrow
\tau_{i+1}\tau_i$ obviously the situation here is more complex. One has the transitions
\begin{eqnarray}
\label{rel1}
10 &\rightarrow 01, & \text{with probability } p,\\
020 &\rightarrow X02, & \text{with probability } p,\\
120 &\rightarrow 210, & \text{with probability } pq,\\
    &\rightarrow 102, & \text{with probability } pq,\\
    &\rightarrow 201, & \text{with probability } p^2,\\
\label{rel6}
121 &\rightarrow 21X, & \text{with probability } p,
\end{eqnarray}
with $X$ being either $0$ or $1$.
In the thermodynamic limit the generating function $\mathcal{M}_2$ for the second-class particle process reads
\begin{equation}
\label{simeq}
\mathcal{M}_2 \simeq \mathcal{M} \frac{p^2\gamma}{q(1-\sqrt{q})^2}, \text{ with } \mathcal{M}=\frac{\gamma}{(1-\gamma)\sqrt{qw_{ud}}}.
\end{equation}
The expression for $\mathcal{M}$ is the corresponding one for the open-boundary TASEP and $\gamma$ is defined in (\ref{def_z}), see Appendix \ref{AppC}.
In the thermodynamic limit, only the denominator $(1-\gamma)\sqrt{qw_{ud}}$ of $\mathcal{M}$ in (\ref{simeq}) is physically relevant. The other factors contribute just a finite number of excursions to the corresponding Motzkin path \cite{Blythe_dyck}. Therefore the second-class particle process has the same thermodynamic physics as the open TASEP. Note that for random update both expressions for $\mathcal{M}$ and $\mathcal{M}_2$ are even the same. This is due to the fact that for random update the weights for $N$ particles on a lattice with $L$ sites are the same for open boundaries and a second-class particle on the ring.\\
Reference \cite{woelki_cargo} shows the equivalence of this process and the one studied in \cite{woelki_par}. It focuses on the thermodynamic limit and simplifies some results of \cite{woelki_par}. The velocity of the 2nd-class particle is shown to be calculated through \cite{woelki_par}
\begin{equation}
\label{2nd_v1}
v=p(1-\rho_+)(1-p\rho_-)-p\rho_-.
\end{equation}
Here $\rho_-$ and $\rho_+$ are the densities directly behind and in
front of the 2nd-class particle $2$. The dynamics of the 2nd-class particle is obtained from
(\ref{rel1}-\ref{rel6}). It moves either forward if it has a hole in
front while at the same time there is no
particle directly behind that simultaneously catches the cargo:
first term in (\ref{2nd_v1}). Or it moves backwards if it has a particle behind:
second term in (\ref{2nd_v1}). With the help of Appendix \ref{asy2} one finds \cite{woelki_cargo}
that the neighboring densities are
\begin{eqnarray}
\label{2nd_dens}
\rho_-=&\dfrac{1}{p}\dfrac{p\rho-J}{1-2J}, \quad\quad\quad
1-\rho_+=&\left( \dfrac{J}{p\rho}\right)^2.
\end{eqnarray}
which yields
\begin{equation}
\label{2nd_v}
\frac{v(\rho)}{p}=\dfrac{1-2\rho}{1-2J}.
\end{equation}
Hence the velocity of the 2nd-class particle vanishes only at half filling $\rho=1/2$. Note that this has the form of a group velocity thus the 2nd-class particle is travelling with the velocity of the density disturbance. This feature should of course be ensured by a 2nd-class particle dynamics and it is a further underpinning of the fact that the model is able to describe the TASEP properly on hydrodynamic scale.
\section{Conclusion}
\begin{figure}
\centering
\includegraphics[height=2.0cm]{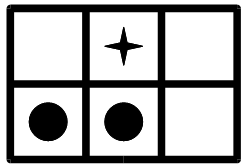}
\caption{Example of a local configuration 120: shown are two chains, the cargo (star) moves in the upper chain; usual particles move in the lower chain. The cargo particle is always carried by a usual particle; therefore it can move independently to the left with probability $p$ (while its carrier either moves to the right ($p^2$) or not ($pq$)); if not, it stays on its carrier with probability $q$ which means that it either moves to the right with its carrier ($pq$) or stays with it on the current site ($q^2$).}
\label{sit120}
\end{figure}
To conclude, we have studied the parallel TASEP with open boundaries where every particle move occurs with probability $p$. This case defines the physics of the maximum-current phase for all $p\in (0,1)$. The main result is the generating function of arbitrary particle numbers $N$ in a system of size $L$ for parallel update. We established an analogy to weighted Motzkin paths and derived an equivalence to ($u,l,d$)-colored Motzkin paths appearing in \cite{Schork}. Already known interpretations of TASEP normalizations with colored Motzkin paths \cite{Blythe_PASEP}, \cite{Duchi} for random-sequential update correspond in this terminology to $(1,2,1)$-colored paths, i.e.\ where only level steps appear in two colors. Here, we found the most natural interpretation as $(2,3,2)$-colored paths. The reason for this increase in the necessary number of colours is the complexity of the parallel update. It is known, that driven-diffusive systems under parallel dynamics typically have a quartic matrix algebra \cite{ERS}, \cite{woelki_par}, \cite{arita} rather than a quadratic one as under random update, ordered-sequential or sublattice parallel update \cite{Blythe}. In fact, we could alternatively use $(2,4,2)$-colored paths for the parallel update so that in contrast to the other updates the number of colours is doubled. This nicely reflects the increase from a quadratic to a quartic algebra. It is expected that the structure found here also holds for similar models with parallel update while the other possible updates mentioned above lead to a ($1,2,1$)-colored path. In the $(2,3,2)$ Motzkin path, red steps are associated with particles or holes while black steps correspond to particle-hole pairs. Therefore paths with the same length do not necessarily correspond to TASEP configurations of the same system size.
One could alternatively represent particle-hole pairs by a succession of two steps, see also Motzkin paths with higher rank considered in \cite{Schork}. However those paths would not end at height $0$, in general. The fact that a fraction of the particle-hole domain walls corresponds to Motzkin steps in a different colour somewhat reflects the fact that under parallel update such a fraction contributes to the flow.
In this paper it was shown how known lattice-path interpretations for the parallel TASEP \cite{Blythe_dyck}, \cite{Blythe} are mapped onto the present one that keeps information of the particle number. Closely related to those problems concerning a given number of particles is the second-class particle process \cite{Derrida_2class}. For parallel update such a process is not as straightforward to define \cite{woelki_par}, though. Here we presented one possible choice of second-class particle dynamics in terms of a cargo process \cite{woelki_cargo} for which the generating function shares the thermodynamics with the generating function that has been derived here for the TASEP.\\
Since the parallel TASEP is equivalent to the Nagel-Schreckenberg model with maximal velocity 1, the results also apply to a bottleneck situation in traffic flow. The average density is $1/2$, however, the density fluctuates around this value. With the help of the generating function obtained in this paper one finds the probability that the road section contains any density of cars. Further investigations shall consider the fluctuations of the particle number as in \cite{Derrida_Brownian} using the findings of the present paper. This is planned to be done in the near future. Additionally it would be interesting to compare the joint current-density distribution with the one for random-sequential update \cite{Blythe}. To finally generalize present results to arbitrary $\alpha$ and $\beta$ one proceeds as in \cite{Mallick}: Denote by $H_{N,M}$ the sum of all possible products of $N$ $D$-matrices and $M$ $E$-matrices (with fugacities $x$ and $y$) in which particle-hole domain-walls are weighted by $q^{-1}$. One observes that $H_{N,M}$ obeys the recursion $H_{N,M}=xH_{N,M-1}D+yH_{N,M-1}E+pq^{-1}xyH_{N-1,M-1}DE$. In order to obtain the weight for $N$ particles and $M$ holes one considers $\langle n|H_{N,M}|m\rangle$ and obtains a recursion that can be solved. Once having $\langle n|H_{N,M}|m\rangle$ one uses the representation of the boundary vectors $\langle W|=\sum_{n\geq 1}((p-\alpha)/\alpha\sqrt{q})^{n-1}\langle n|$, $|V\rangle=\sum_{m\geq 1}((p-\beta)/\beta\sqrt{q})^{m-1}| m\rangle$ to find $\langle W|H_{N,M}|V\rangle$. The result is no longer a Motzkin path but a sum over lattice paths. Walks that start at height $n$ and end at height $m$ each contain an additional factor $((p-\alpha)/\alpha\sqrt{q})^{n-1}((p-\beta)/\beta\sqrt{q})^{m-1}$.
\appendix
\section{Preliminaries for the parallel TASEP}
\label{prelim}
In \cite{woelki_par} a slightly different model for fixed particle number was considered but results can be used here. Application of those results (compare equation (A.2) in \cite{woelki_par}) to the TASEP yields
\begin{equation}
\label{ZS}
\mathcal{Z}_L(x,y)=S_L(x,y)+px S_{L-1}(x,y), \quad {\rm with}\; S_L=m_L + py m_{L-1} \text{ and } m_L=\langle 1|G_L|1 \rangle.
\end{equation}
 Note that the $S_L(x,y)$ recover the $z_L$ appearing in (\ref{norm}) for $x=y=1$ and therewith $\mathcal{Z}_L$ recovers $Z_L$. The matrices $G_n$ are defined recursively through the matrices $C=xD+yE$ and $K=q^{-1}DE=D+E+p$, namely $G_n=G_{n-1}C+pxyG_{n-2}K$ with the initial condition $G_{-1}=0$ and $G_0=1$. Iterating this equation yields
\begin{eqnarray}
G_0(x,y)&=&1,\nonumber\\
G_1(x,y)&=&C,\nonumber\\
G_2(x,y)&=&C^2+pxyK,\nonumber\\
G_3(x,y)&=&C^3+pxy(CK+KC),\nonumber\\
G_4(x,y)&=&C^4+pxy(C^2K+CKC+KC^2)+(pxy)^2K^2,\nonumber\\
\label{GListe}
G_5(x,y)&=&C^5+pxy(C^3K+C^2KC+CKC^2+KC^3)+(pxy)^2(CK^2+KCK+K^2C)\\
\dots &=& \dots\nonumber
\end{eqnarray}
Eventually taking the matrix element one is drawn to
\begin{equation}
\label{mncompl}
m_L:=\langle 1 | G_L | 1 \rangle = \sum\limits_{B=0}^{L/2}(pxy)^B f_{L-B,B},
\end{equation}
where we
defined
\begin{equation}
\label{def_f}
f_{L-B,B}=\langle 1|\prod\limits_{i=0}^{L-B}\sum\limits_{\sigma_i=0,1}\left(C\delta_{\sigma_i,0}+K\delta_{\sigma_i,1} \right) \cdot\delta_{\sum_{\sigma_i},B} |1 \rangle.
\end{equation}
The term in the parenthesis denotes the sum over all matrix products of $C=xD+yE$ and $K=D+E+p$ with exactly $B$ $K$s and $L-2B$ $C$s so that $B=0$ recovers $\langle 1|C^L|1 \rangle$. The generating function becomes with (\ref{mncompl})
\begin{equation}
\label{Mf}
\mathcal{M}(x,y,\lambda,\mu)=\sum\limits_L m_L \lambda^L = \sum\limits_{L}\sum\limits_{B=0}^{L/2}(pxy)^B f_{L-B,B}\mu^{L-B}\lambda^L.
\end{equation}
To obtain (\ref{Mxy}), (compare equation (A.2) in \cite{woelki_par}) one changes the summation. Instead of summing (\ref{GListe}) line-wise and taking the matrix element as in (\ref{Mf}), one sums diagonal-wise, so to speak: multiply (\ref{GListe}) by $\lambda^n$ and sort the results depending on the lengths of words in $C$ and $K$ that occur to arrive at the matrix element
\begin{equation}
\label{ab}
\mathcal{M}(x,y,\lambda,\mu)=\sum\limits_{n}\mu^{n}\sum\limits_{b=0}^{n}(pxy)^b f_{n,b}\lambda^{n+b}.
\end{equation}

\section{Results for the second-class process}
\label{asy2}
In \cite{woelki_cargo} it was argued that the second-class particle process can be mapped onto the (single species) traffic model in \cite{woelki_par} which is defined by the transitions $100\rightarrow 001$ and $101\rightarrow 011$. We repeat shortly the arguments here. If the total number of holes in the traffic model is an even number, then in the stationary state every particle $\mu$ has an even number of holes $2n_\mu$ in front. Thus by mapping $00\rightarrow 0$ one has $2n_\mu \rightarrow n_\mu$ and the usual TASEP is recovered. If the total number of holes is an odd number, the dynamics tries to do the same, however one odd-valued headway remains. The stationary configurations are of the form $\{2n_1, 2n_2, \dots, 2n_N+1\}$. If one changes the numbering of particles after every update, one is able to achieve that the first particle always has the single 'excess hole' behind. Now one defines the second-class particle $2$ as the unit $01$, i.e.\ the excess hole plus the following particle. The rules (\ref{rel1}-\ref{rel6}) then are just a consequence of the simultaneous updating. The normalization found in \cite{woelki_par} is
\begin{eqnarray}
\mathcal{Z}_{p,\beta}(x,y,\lambda)=\frac{\gamma_-(1-\gamma_-)}{x\lambda (1-s)}\cdot
\frac{1}{1-\gamma_{-}\left(1+\frac{p-\beta}{\beta(1-p)}(1-s)^{-1}\right)}
\cdot\left[\frac{1+py\lambda}{1-\gamma_{-}}-\frac{\beta}{1-\gamma_- s}\right]
\end{eqnarray}
which simplifies for $p=\beta$ to
\begin{eqnarray}
\label{eq:Z}
\mathcal{Z}_{p,p}(x,y,\lambda)=\frac{\gamma_-}{x\lambda (1-s)}\cdot\left[\frac{1+py\lambda}{1-\gamma_{-}}-\frac{p}{1-\gamma_- s}\right]
\end{eqnarray}
with
\begin{equation}
s=s(p,x,y,\lambda)=1-\sqrt{\frac{y+pxy\lambda}{x+pxy\lambda}} = 1-\sqrt{w_d/w_u}
\end{equation}
and $\gamma_-$ is one of the roots of the equation
\begin{equation}
\label{def_z}
\frac{\gamma(1-\gamma)}{q+p\gamma}=z.
\end{equation}
Note that rewriting this expression as
\begin{equation}
\label{peak}
\gamma= qz + z\frac{\gamma}{1-\gamma}
\end{equation}
shows that $\gamma$ is in fact the generating function for the number of weighted-peak walks in \cite{Blythe_dyck}.
Those are excursions on the rotated square lattice with a fugacity $qz$ for each peak and a fugacity $z$ for each up-down pair so that in total $z$ is conjugate to the length of the excursion. The parameter $z$ in (\ref{def_z}) is in fact a function $z=z(p,x,y,\lambda)$:
\begin{equation}
z=\frac{\lambda\sqrt{(x+pxy\lambda)(y+pxy\lambda)}}{1-\lambda\left[\sqrt{(x+pxy\lambda)(y+pxy\lambda)}(p-2)+(1-p)(x+y+2pxy\lambda)+p^2xy\lambda\right]}
\end{equation}
that is rewritten with (\ref{lud}) as
\begin{equation}
\label{zw}
z = \frac{\sqrt{w_{ud}/q}}{1 - w_l + (1+q) \sqrt{w_{ud}/q}}.
\end{equation}
In the thermodynamic limit the process is dominated by the square-root singularity in $\gamma$. This can be written as a condition for $z$:
\begin{equation}
\label{zsimeq}
z\simeq \frac{(1-\sqrt{q})^2}{p^2}.
\end{equation}
The asymptotics of the model can be obtained from the grand-canonical partition function
\begin{equation}
\label{Zasymp}
\mathcal{Z}_L\sim(1+px\lambda)\lambda^{-L}.
\end{equation}
In (\ref{Zasymp}), $x$ and $\lambda$ are the thermodynamic particle and lattice fugacities that were first found in \cite{woelki_par}. They can further be simplified to \cite{woelki_cargo}
\begin{eqnarray}
\label{2nd_fug}
x=\left(\dfrac{(\rho-J)(p\rho-J)}{(1-p)\rho J}\right)^2, \quad\quad\quad
\lambda=\dfrac{(1-p)(1-\rho)^2}{[1-J-p(1-\rho)]^2},
\end{eqnarray}
with the density $\rho=N/L$ in the second-class particle process and $J(\rho)=(1-\sqrt{1-4p\rho(1-\rho)})/2$ the current-density relation. The identity of (\ref{2nd_fug}) and the expressions given in \cite{woelki_par} can be verified most easily using a computer algebra system.

\section{TASEP versus 2nd-class particle process}
\label{AppC}
\label{Eq}
In this appendix the generating functions $\mathcal{M}_2$ and $\mathcal{M}$ of the second-class particle process and the TASEP respectively are derived that share the same thermodynamics.
For simplicity we take $\lambda=1$ without loss of generality as explained in the main text. To calculate
\begin{equation}
\label{F}
(1-\tilde{C}(x,y))^{-1}=\sum\limits_{n=0}^{\infty}\left(C+pxy K \right)^n,
\end{equation}
the idea in \cite{woelki_par} is to transform the matrices similar to \cite{Blythe}. Define
  \begin{eqnarray}
  \label{primed}
D&=&\sqrt{w_d/w_u}(D'-q)  + q, \quad \quad E=\sqrt{w_u/w_d}(E'-q)  + q.
  \end{eqnarray}
  One can check that this transformation leaves the algebra invariant, i.e.\
  \begin{equation}
  \label{DEprimed}
  D'E'=q(D'+E'+p)=qK', \quad \langle W|E'=q\langle W|, \quad D'|V\rangle = q|V\rangle.
\end{equation}
In this notation $C+pxy K$ becomes finally
\begin{eqnarray}
\tilde{C}=C+pxy K=\sqrt{w_{ud}/q}K'+w_l-(1+q)\sqrt{w_{ud}/q}.
\end{eqnarray}
It can be checked that the Motzkin-path transfer matrix $\tilde{C}$ is thus changed in a way so that (while level steps have same weight $w_l'=w_l$) one finds for the weights of up-steps and down-steps $w_{u}'/w_u=\sqrt{w_u/w_d}$ and $w_{d}'/w_d=\sqrt{w_d/w_u}$, respectively so that $w_{ud}'=w_{ud}$.
One now executes the sum in (\ref{F}) to find with $z$ from (\ref{zw}) that
\begin{equation}
(1-\tilde{C})^{-1}= (1-w_l+(1+q)\sqrt{w_{ud}/q})^{-1}(1 - zK')^{-1}.
\end{equation}
One proceeds by using $K'=D'+E'+p$. With the help of $\gamma$ being determined by (\ref{peak}) one writes
\begin{equation}
(1 - zK')^{-1}=\left(1-\gamma/qE' \right)^{-1}\left(1-\gamma/ qD' \right)^{-1}\frac{q+p\gamma}{q},
\end{equation}
so that
\begin{equation}
\label{F_end}
(1-\tilde{C})^{-1}=\frac{q+p\gamma}{q}\frac{1}{1-w_l+(1+q)\sqrt{w_{ud}/q}}\left(1-\gamma/qE' \right)^{-1}\left(1-\gamma/q D' \right)^{-1}.
\end{equation}
Using (\ref{DEprimed}) the matrix element of $(1-\tilde{C})^{-1}$ can be taken that recovers $\mathcal{M}$:
\begin{equation}
\label{erzM}
\mathcal{M}=\frac{q+p\gamma}{q}\frac{1}{1-w_l+(1+q)\sqrt{w_{ud}/q}}\frac{1}{(1-\gamma)^2}
\end{equation}
which can be written with (\ref{def_z}) and (\ref{zw}) as
\begin{equation}
\mathcal{M}=\frac{\gamma}{(1-\gamma)\sqrt{qw_{ud}}},
\end{equation}
and finally as
\begin{equation}
\label{erzMshort}
\mathcal{M}=\left(\frac{\gamma}{z}-q\right)\frac{1}{\sqrt{qw_{ud}}}.
\end{equation}
This is compared to the second-class particle process. Its algebra differs slightly from the TASEP. Instead of $\langle 1|E = q\langle 1|$ one has \cite{woelki_cargo} $\langle 1 | EE = q\langle 1 | E$ and $\langle 1 | ED = q\langle 1 | (D+p)$. Therefore we find from (\ref{F_end}) a slightly different result that can be written as
\begin{equation}
\label{erz2}
\mathcal{M}_2=\mathcal{M}\frac{q+p\gamma}{q(1-\gamma)}.
\end{equation}
Now using (\ref{peak}) and the asymptotic expression (\ref{zsimeq}) gives (\ref{simeq}).

We now prove the identity of (\ref{Mxy2}) and (\ref{erzMshort}) involving the function $\gamma(q,z)$ that appears in \cite{Blythe_dyck} in several lattice-path interpretations for the parallel TASEP, see Appendix \ref{asy2}. First, note that (\ref{peak}) can be rearranged to $\gamma(1-pz-\gamma)=qz$. Now solving (\ref{erzMshort}) for $\gamma$ and inserting the resulting expression yields $\mathcal{M}=1+\tilde{w}_l\mathcal{M}+w_{ud}\mathcal{M}^2$ with $\tilde{w}_l=1+(z+qz-1)\sqrt{qw_{ud}}/(qz)$. Finally using (\ref{zw}) leads to $\tilde{w}_l=w_l$ which completes the proof.

\end{document}